# Soft Lithography using Nectar Droplets

Saheli Biswas[1], Aditi Chakrabarti[1], Antoine Chateauminois[2], Elie Wandersman[3], Alexis M. Prevost[3], Manoj K. Chaudhury[1]*

[1]Department of Chemical and Biomolecular Engineering, Lehigh University, Bethlehem Pennsylvania 18015 USA

[2]Soft Matter Science and Engineering Laboratory (SIMM), CNRS/UPMC Univ Paris 6,UMR 7615, ESPCI, F-75005 Paris, France

[3]CNRS, UMR 8237, Laboratoire Jean Perrin (LJP), F-75005, Paris, France
Sorbonne Universités, UPMC Univ Paris 06, UMR 8237, Laboratoire Jean Perrin, F-75005, Paris, France

**Abstract.** Inspite of significant advances in replication technologies, methods to produce well-defined three dimensional structures are still at its infancy. Such a limitation would be evident if we were to produce a large array of simple and, especially, compound convex lenses, also guaranteeing that their surfaces would be molecularly smooth.  Here, we report a novel method to produce such structures by cloning the 3D shape of nectar drops, found widely in nature, using conventional soft lithography.The elementary process involves transfer of a thin patch of the sugar solution coated on a glass slide onto a hydrophobic substrate on which this patch evolves into a microdroplet. Upon the absorption of water vapor, such a microdroplet grows linearly with time and its final size can be controlled by varying its exposure time to water vapor. At any stage of the evolution of the size of the drop, its shape can be cloned onto a soft elastomer by following the well-known methods of molding and crosslinking the same. A unique new science that emerges in our attempt to understand the transfer of the sugar patch and its evolution to a spherical drop is the elucidation of the mechanics underlying the contact of a deformable sphere against a solid support intervening a thin liquid film. A unique aspect of this work is to demonstrate that higher level structures can also be generated by transferring even smaller nucleation sites on the surface of the primary lenses and then allowing them to grow by absorption of water vapor. What results at the end is either a well-controlled distribution of smooth hemispherical lenses or compound structures that could have potential applications in the fundamental studies of contact mechanics, wettability and even in optics.

1.Introduction

Recently, there has been a significant interest in producing model rough surfaces to enable fundamental studies in wetting, contact mechanics, and tribology.[1-3] Such studies require a detailed knowledge of the uniformity, spacing and distribution of the size of the surface asperities, and their interactions with a proximal surface. Several methods to produce microlenses have already been reported in the literature.[4-18] Among these, a simple and inexpensive technique uses a micro-milling procedure that results in either uniform or random spatial distribution of hemispherical depressions on a rigid polymer.[6,7] Once such a surface is produced, its negative can be prepared by crosslinking a PDMS (Polydimethylsiloxane) elastomer against the concave depressions of the master. While the micro-milling technique provides by design an accurate control over the type of order (or disorder) needed to produce model rough surfaces, the resulting hemispheres have microscopic and/or sub-microscopic corrugation on top of the primary asperities, which complicates its systematic use in various tribological studies in that the micro-asperity contacts cannot be assimilated to single asperity contacts. Smooth convex asperities can, however, be produced either by the process of self-assembly of microroplets in the form of breath figures, or via a sequential production of a negative and then a positive image of microscopic liquid droplets condensed onto a hydrophobic surface.[6] While the concave hemispherical asperities prepared in this manner are very smooth, controlled distribution of the size and spacing of the asperities cannot be easily achieved using this method due to the intrinsic Ostwald ripening related coarsening of the drop size distribution.

What would be ideal is a synergy of the above two methods that would offer the flexibility of the control of asperity size and their spatial distribution by micro-milling on one hand, and the high degree of smoothness achieved with liquid droplet based lithography on the other. Here we report a new method for producing such an array of controllable distribution of super-hemispherical elastomeric lenses by cloning the 3d shapes of nectar droplets having specific size and predetermined distribution of its size. This unique methodology takes advantage of *directed selective growth of the droplets on specified locations* on a hydrophobic surface using strongly hydrophilic nucleation sites that avoids the usual uncontrolled growth of droplets via condensation at random locations on the surface. Apart from being simple, fast and inexpensive, one unique advantage of this method is that a secondary, and even a tertiary level of structures can be generated above the first generation of structures thus producing compound lenses that have potential applications in optics as well as contact mechanics studies.

In order to achieve this objective, we mimic Nature's way of controlling droplet growth and preserving its high water content in varieties of situations.[19, 20] All naturally occurring nectars, *e.g.* Agave, Honey or Maple syrup, are rich aqueous solutions of dextrose such as glucose and fructose that do not crystallize even at high concentrations of the sugars.[17] As the chemical potentials of water in these nectar solutions are significantly lower than that of the ambient, it thwarts its rapid drying out as well. The proposed method is also coupled to a decades old interesting discovery by Williams et al.[21], who noted that while a salt crystal placed on a hydrophobic surface grows by absorbing water vapor, a band of dry zone is created around it. A similar finding was reported later by Leach et al.[22] as well. Following these leads, we observed through the course of our own work that when such hygroscopic droplets of concentrated sugar solutions are close to each other, the dry bands around them are superimposed in such a way that no growth of droplet occurs in their intervening spaces. This observation, combined with the fact that the kinetics of the droplet growth and its limiting size can be manipulated by controlling the surrounding humidity and the initial volume of the seeding droplet, allowed us to develop an ultra-soft lithographic process in which the arrangements and the shapes of the droplets could be image wise replicated onto a soft elastomer.

In what follows next, we first report the formation of droplet depleted bands on a hydrophobic substrate (a polystyrene Petri dish) using the precursors of a concentrated aqueous solution of glucose and fructose as hygroscopic sites for water vapor condensation. This was accompanied by a study of the kinetics of growth of a single droplet of the sugar solution under humid condition in order to estimate the time of exposure of the hygroscopic sites to water vapor for a desired replication process. In order to generate an array of hemispherical drops, a PDMS elastomer consisting of the convex hemispherical surface protrusions was stamped against a spin coated thin film of the sugar solution on a glass slide. Transfer of the droplet footprints onto a polystyrene Petri dish and its subsequent exposure to water vapor afforded formation of a replica of the droplet pattern. In order to understand the mechanics of the transfer of a patch of sugar solution supported as a thin film on a glass slide, we carry out the contact mechanics experiments in which the area of contact formed by PDMS lenses of various radii against the solution coated glass slides at different loads was examined that allowed us to establish how the footprint of the final daughter lens is related to that of the parent lens. The three dimensional morphology of this replica was then transferred to a PDMS elastomer using conventional soft lithography.[6,23-26] We conclude by illustrating how the above methods could also be used to produce secondary level structures on the first generation of lenses, thus producing compound lenses.

**2. Condensation of water vapor on a single syrup droplet**

We first duplicated the report in literature that a dry band forms around a hygroscopic site exposed to water vapor, albeit with a sugar solution (*See Section 9.1, Experimental details*), not a salt solution.[21,22] The aqueous solution of glucose and sucrose affords the unique advantage that a supersaturated solution of sucrose can be prepared, the crystallization of which is inhibited by glucose.[27] A small (0.04μL) droplet of sugar solution was deposited on the inner side of the lid of a polystyrene Petri dish, which was exposed to water vapor saturated air (relative humidity ~ 99.99%). Humidity was controlled by placing warm water (35°C) in the base of the Petri dish and covering it with the lid carrying the sugar droplet. It was observed under a microscope that as the water vapor was absorbed by the sugar droplet, the latter grew in size. A 0.4mm thick band, however, developed around the droplet with no visible condensation of water vapor (Figure1A) in that zone. The thickness of the band increased with time as well, although in this work we have not analyzed the growth rate of the band, either as a function of time or initial size and composition of the drop. We often refer to this band as the depletion zone in this paper. No such depletion zone was observed with a pure water droplet (0.3μL) (Figure1B) although a very thin (0.02mm) dry zone developed around the growing water droplet, which seems to result from the imminent fusion of the small drops with the large one.

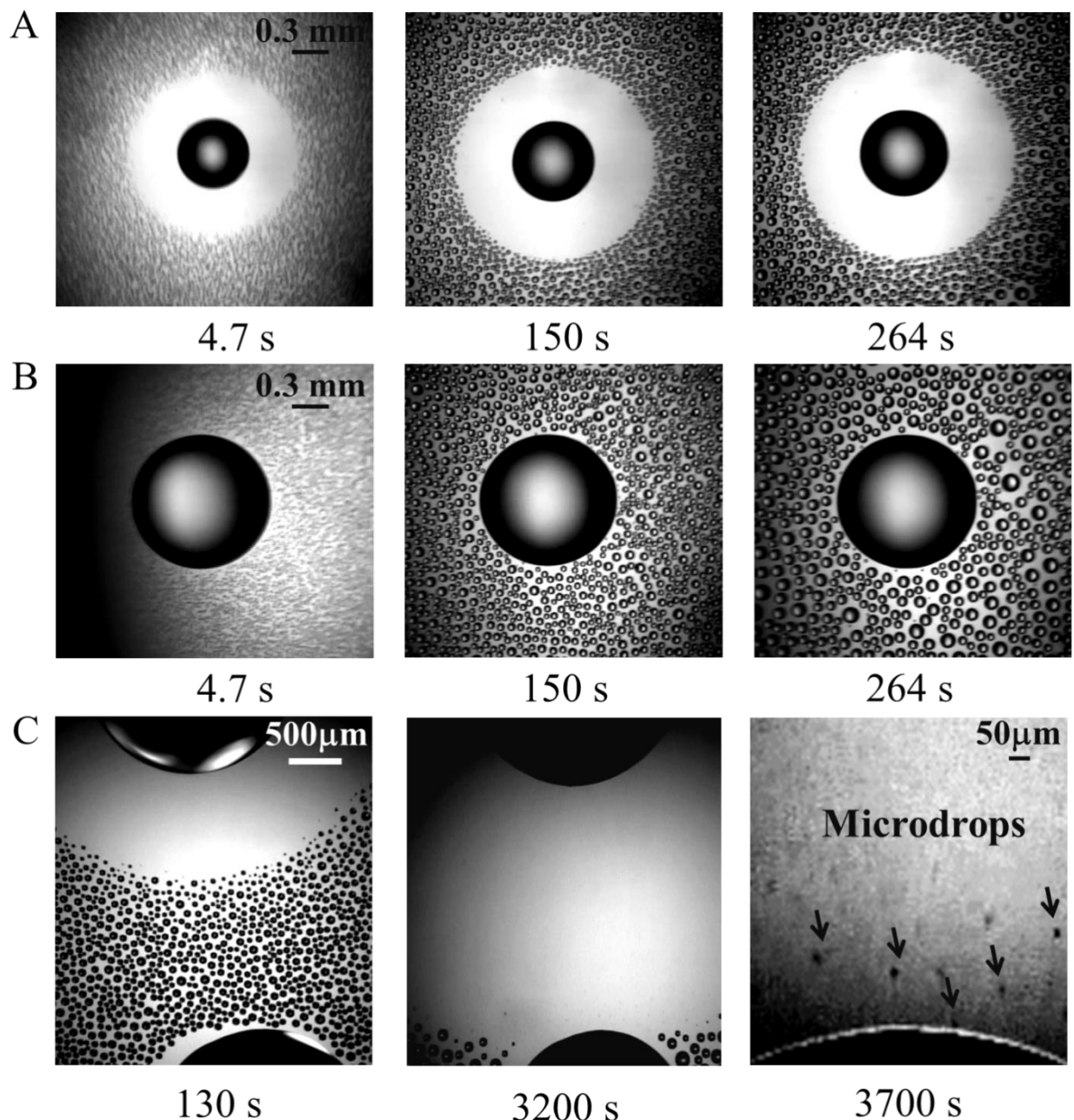

**Figure1.** (A) Formation of a dry band around a droplet of syrup (0.04μL) when exposed to saturated humid atmosphere. (B) In a similar experiment with a pure water droplet (0.3μL), only a very thin band devoid of micro-droplets is evident. (C) A sugar drop (upper) and a water drop (lower), both having a volume ~ 0.7μL, deposited in close proximity on a hydrophobic surface covered with pre-condensed water droplets. Evaporation of the condensed drops close to the sugar drop generates a dry band around it that grows with time. The dry band eventually reaches the water drop by means of the evaporation of all the condensed water droplets from the intervening region. Traces of micro-droplets near the water drop are pointed by the arrows in the last frame of the bottom panel, which has been magnified by a factor of 4 to enhance their visibilities.

Our findings are in agreement with the previous observations reported by Williams et al.[21] and Leach et al.[22], who observed that a hygroscopic drop grows with the concomitant formation of a dry annular zone around its boundary owing to the lowering of vapor pressure in the regions near the drop. These authors suggested that the difference between the partial pressures of the surrounding vapor and that of the drop is the driving force for vapor transport and its absorption by the hygroscopic drop. This flux of vapor is higher with a sugar drop than that with a water drop, where there is very small driving force for water vapor condensation. The increase in net flux creates a concentration gradient in the vapor around the sugar drop, which results in the formation of the nucleation free annular band as the vapor pressure there is not large enough to allow condensation on the surface.

This finding is consistent with another new experiment, in which small droplets were condensed on a polystyrene surface prior to depositing a droplet of aqueous sugar solution onto it. In this case, the condensed droplets near the sugar drop continue to evaporate in the absence or presence of another water drop placed near it (Figure1C). As the dry zone increases in size, a few droplets close to the water drop shrink, but do not evaporate completely; very small drops remain visible near the water drop for a very long time. They vanish only when the water droplet continues to shrink, and finally, disappear. The stabilization of these tiny droplets near the water drop may suggest that a gradient of even smaller stable micro-droplets may be present on the surface that are not detected with our optical microscope. This interesting effect could perhaps be interrogated with an environmental electron microscope that promises to unfold the stabilization of micro- or nano- drops on hydrophobic surfaces.

## 3. Kinetics of growth and decay of a syrup droplet

Just as a drop grows in size when exposed to water vapor, it shrinks when exposed to a dry atmosphere. We examined the kinetics of the growth and the decay of the size of a syrup drop by placing a small (0.18μL) droplet on the tip of a glass bead that was formed by melting a fine glass capillary tube and silanizing it with Perfluorooctyltrichlorosilane. The small droplet of the sugar solution was deposited in the glass bead by bringing it in contact with a thin film (18μm thickness) of sugar solution spin-coated on a glass slide, followed by quick removal. This glass bead with the syrup drop was then placed inside a glass chamber saturated with water vapor, the growth of which was observed under a microscope. The drop indeed grew (Figure2A) much like that of a Buller's drop that forms at the hilar appendage of a basidiospore prior to its forced discharge[28,29] resulting from the coalescence of two water droplets. After the droplet grew to about 1.5μL, it was exposed to dry air (humidity~ 0%) that resulted in the evaporation of water and thus the shrinkage of the drop, ultimately reaching its original size (Figure2B). Both the growth and decay processes were imaged using a CCD camera and subsequently analyzed using an image processing software (*see Section 8.1, Experimental details*). It is remarkable that the drop volume increased linearly with time in the growth phase (Figure2C), whereas it decayed following a power law (Figure2D) during the shrinkage phase. The following simple analysis suggests a possible explanation for the difference between the growth and decay kinetics. We write down an approximate equation of the kinetics of the droplet growth, where the driving force arises from the difference of the partial pressure of the vapor ($P^*$) far from the drop and that ($P$) just outside of it:

$$\frac{dN_w}{dt} = kA[P^* - P] \tag{1}$$

where, $N_w$ denotes the moles of water in the drop, $k$ is the mass transfer coefficient for diffusion of water vapor from the bulk vapor phase to the surface of the drop and $A$ is its surface area. During the desorption phase, the driving force originates from the difference between the partial vapor pressure just outside the drop and that ($P^*=0$) of the dry air:

$$-\frac{dN_w}{dt} = kAP \tag{2}$$

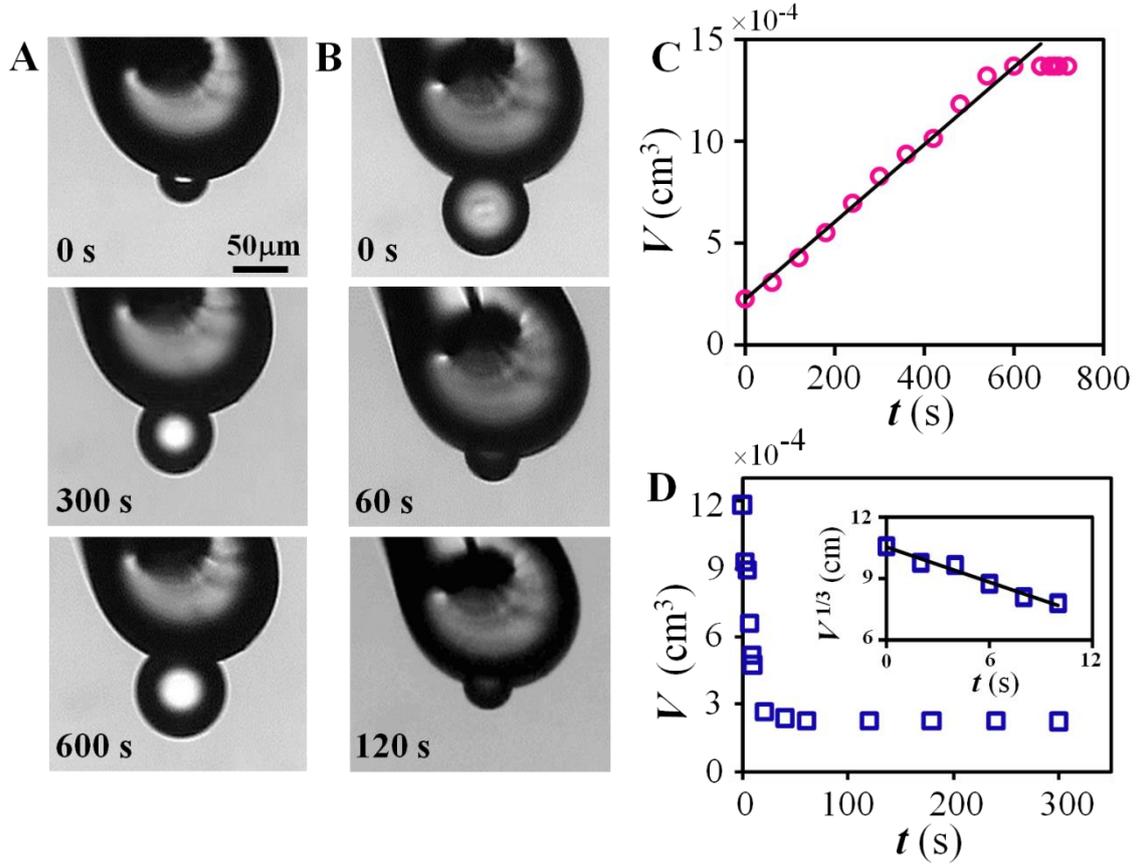

**Figure 2.** (A) Growth of a small droplet of the sugar solution adhering to the tip of a hydrophobic glass bead in a saturated humid atmosphere. (B) Shrinkage of the fully grown sugar droplet in dry air. (C) The pink open symbols represent the volume of the sugar drop increasing linearly as a function of time that has been fitted by the black solid line. (D) The non-linear temporal evolution of the volume of the drop (blue squares) in dry air. The inset is a linear plot of $V^{1/3}$ against $t$ fitted with a black solid line.

Using appropriate vapor-liquid equilibrium and mass balances, Equation 1 and 2 can also be reduced as follows:

$$\frac{dN_w}{dt} = (36\pi)^{1/3} k[P^*(N_w\bar{v}_w + N_s\bar{v}_s)^{2/3} - HN_w(N_w\bar{v}_w + N_s\bar{v}_s)^{-1/3}] \tag{3}$$

$$-\frac{dN_w}{dt} = kHN_w(N_w\bar{v}_w + N_s\bar{v}_s)^{-1/3} \tag{4}$$

where $N_s$, $N_w$ denote the number of moles of sugar and water in the syrup drop respectively and, $\bar{V}_s, \bar{V}_w$ represent their molar volumes which are related to the total volume $V$ of the drop as $V = N_w \bar{V}_w + N_s \bar{V}_s$. $H$ is Henry's law constant that is given as $H=P/C_w$, $C_w$ being the concentration of water in the solution (*See Appendix A2*).

In order to gain an insight into the absorption kinetics, let us consider the short time limit of Equation 3, in which the first term on the right hand side of the equation dominates and $N_w \ll N_s$. The solution of the resulting trivial equation yields a linear growth of $N_w$, and thus the volume of the drop, with time $t$:

$$N_w \sim kP^*(N_s \bar{V}_s)^{2/3} t \tag{5}$$

On the other hand, at short times of the desorption phase corresponding to $N_s \ll N_w$, the drying of the drop can be described as:

$$N_w^{1/3} - N_{wo}^{1/3} \sim -kH(\bar{V}_w)^{-1/3} t \tag{6}$$

Thus, the moles of water in the drop and hence the drop volume follows a power law with respect to time during the decay phase, finally reaching to a constant value when it completely dries up.

The differences in the kinetics of the growth and the decay of the drop size as observed experimentally (Figure2) seem to obey the simple kinetic processes described as above. We note, however, that the observed kinetics conforms to the above expectations for times even longer than the limiting cases described here. We believe that this agreement at long times could be fortuitous as the assumptions are strictly valid for short time kinetics. We postpone a detailed analysis of the problem for future, as the main objective of this part of our work is to identify an approximate timescale to allow the droplet growth for the replication processes to be discussed below.

**4. Genesis of a single hygroscopic nucleation site**

In this section, we explore the underlying physics that controls the size of the nucleation site that is the basis for droplet growth and the subsequent lithographic processes. In order to study this subject systematically, it was necessary to employ smooth lenses of PDMS of different radii of curvatures (*See Section 9.2, Experimental details*). To begin with, one of these lenses was brought into contact with, and then separated from, a clean glass slide coated with a thin film of sugar solution that resulted in the transfer of a small amount of solution onto the lens (Figure3A). This lens was subsequently pressed against and released from a polystyrene Petri dish that resulted in the desired transfer of a small hygroscopic precursor onto the Petri dish. Once this surface was exposed to water vapor for 5 minutes, a small drop bloomed onto the surface somewhat like that described in the above section. While the drop grew in size, its base area remained more or less constant within the observation time scale of 5 minutes. Analysis of the image of the hygroscopic precursor revealed that its base radius ($\rho$) varies systematically with the radius of curvature of the parent lens ($R$) following a power law (Figure3D): $\rho \sim R^{0.6}$.

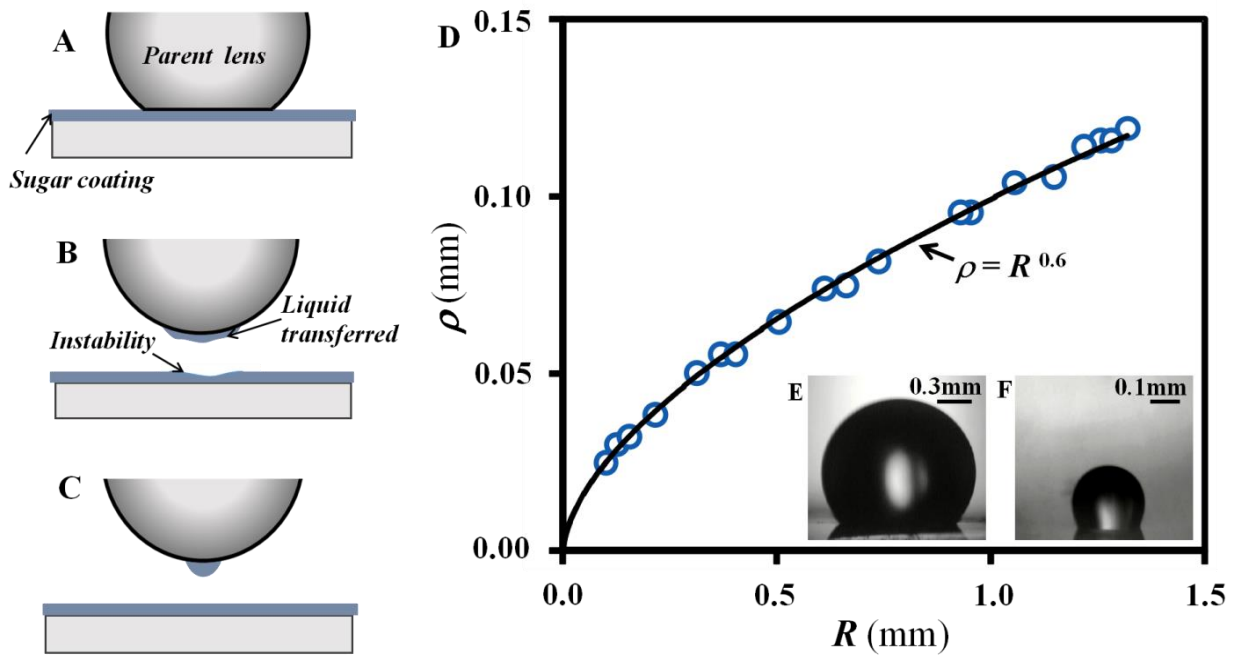

**Figure 3.** (A) A hemispherical PDMS lens is brought into contact with a glass slide coated with a thin film of sugar solution. (B) On pulling off the lens, we envisage that it separates from the glass slide by undergoing a contact mechanical instability. The surface of the transferred liquid eventually smoothens (C). This liquid coated lens is then pressed against and separated from a polystyrene Petri dish, which is also subjected to the similar contact instability as depicted in (B). (D) The base radius ($\rho$) of the syrup footprint increases with the radius ($R$) of the parent lens following a power law. Comparison of the size of the parent super hemispherical lens shown in (E) and the resulting daughter lens shown in (F) obtained by the sequential lithographic processes described in the text.

**5. Mechanics underlying the contact printing of the nectar droplet:** There are two basic steps involved in the contact printing of a hydrophobic surface with a drop of nectar. The first step involves bringing the PDMS lens into contact with a nectar film coated glass slide followed by rapidly separating the two. Small amount of liquid that is transferred onto the lens as a flat patch dewets the lens and forms a hemisphere. In the next step, this lens is pressed against a hydrophobic substrate at a given load that squeezes the drop against the substrate, which subsequently splits into two halves during the retraction of the lens from the substrate. The patch remaining on the hydrophobic substrate then dewets to form a smaller nectar drop, which is then cloned onto PDMS by swelling in water vapor and then using the conventional molding technique using crosslinkable PDMS. All these experiments were performed under ambient conditions (21°C, 35% RH).

When a PDMS lens is brought into contact with a nectar film coated glass slide in the first step, the film is squeezed, but it is not completely displaced from the contact formed by the deformed sphere and the glass slide. An outline of the deformed lens surrounded by a liquid ring can be easily observed under a low power microscope (40X magnification). Both the diameters of the inner and the outer rings decrease with the reduction of the applied load. As a thin liquid film always separates the lens from the glass substrate, the adhesion occurs between the lens and the liquid film in the zone of contact surrounded by a liquid bridge that forms at the meniscus of contact. As a negative Laplace pressure acts in the outer region of contact, the resulting adhesion leads to a JKR (Johnson-Kendall-Roberts) like deformation of the lens[30,31], in which the radius of contact ($a_i$) is related to the applied load ($P$), the work of adhesion ($W$) and the radius ($R$) of the PDMS hemisphere as in equation 7.

$$\frac{a_i^{3/2}}{R} = \frac{P}{Ka_i^{3/2}} + \left(\frac{6\pi W}{K}\right)^{1/2} \tag{7}$$

Where $K$ is an elastic constant given by $K = (4/3)E/(1-\nu^2)$ where $E$ is the Young's modulus and $\nu$ the Poisson's ratio. According to equation 7, a plot of $a_i^{1.5}/R$ versus $P/a_i^{1.5}$ should be linear with the slope and intercept as $1/K$ and $(6\pi W/K)^{1/2}$, respectively. The representative data obtained with five different lenses (Figure 4) show that such a plot is indeed linear for each lens with a positive intercept in the y-axis as is the case with a non-zero adhesion. The slope of the line obtained from such a plot yields the typical elastic constant of PDMS as 2.7MPa and the average work of adhesion as 185mJ/m$^2$, which is on the order of twice the surface tension of water that is consistent with the picture that the adhesion is related to the Laplace pressure[32] acting in the open surfaces of the crack.

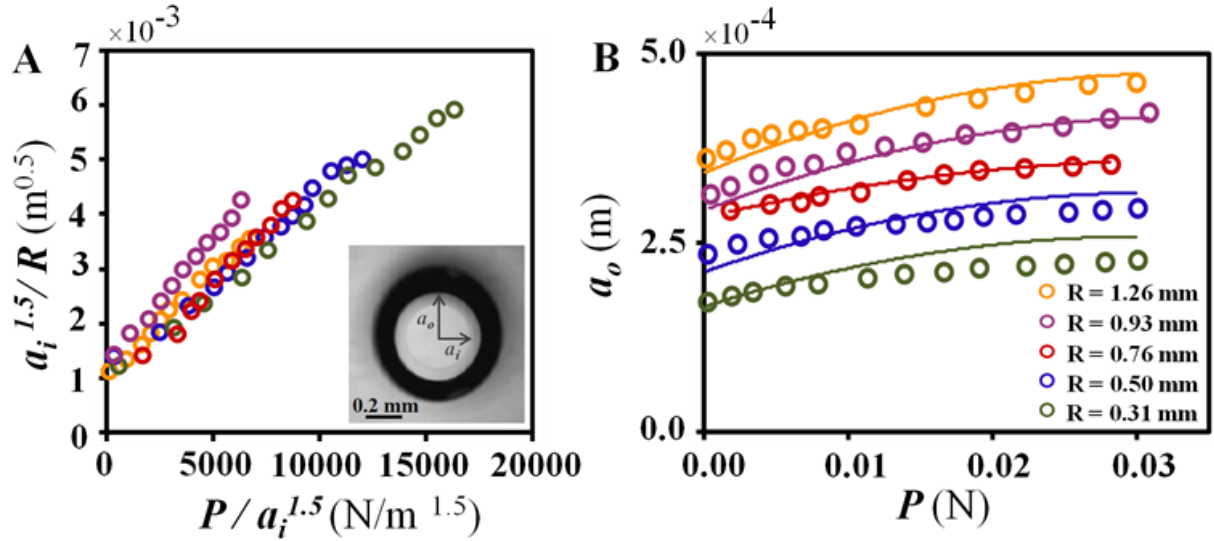

**Figure 4.** (A) JKR-like contact mechanics demonstrated by the different PDMS lenses pressed against a syrup coated glass slide during the unloading cycle. As shown in the inset, $a_i$ denotes the contact radius of the region of the lens in contact with the syrup coated glass slide, undergoing elastic deformation, and $a_o$ denotes the outer contact radius of the syrup on the lens under a load of ~ 34mN. (B) The contact radius $a_o$ of the syrup on the lens plotted as a function of the load $P$ for lenses of different sizes. Solid lines correspond to the theoretical prediction (eqns 7 and 8).

| $R\,(\times 10^3\,\text{m})$ | $K\,(\text{MPa})$ | $W\,(\text{mJ/m}^2)$ |
|---|---|---|
| 0.31 | 3.2 | 210 |
| 0.50 | 3.0 | 192 |
| 0.76 | 2.4 | 152 |
| 0.93 | 2.2 | 199 |
| 1.26 | 2.6 | 169 |

**Table 1.** The elastic constant $K$ and work of adhesion $W$ as calculated from the slope and intercept of the unloading curves for the different PDMS lenses from a syrup coated glass slide as shown in Figure 4A.

The radius of the outer ring depends on the radius of curvature of the lens following a power law $R^{0.5}$ (*See Appendix A1*), which has a finite value for $P=0$ when the sphere makes almost a point contact with the underneath substrate. Taking the thickness of the outer periphery of the liquid bridge to be $\delta$, its radius can be estimated to be $\sqrt{2R\delta}$. We assume that width of the liquid bridge given by the preceding equation remains constant while the lens makes an inner contact governed by the JKR contact mechanics. Thus, we may write

$$a_o \approx a_i + \sqrt{2R\delta} \tag{8}$$

By taking this value of $\delta$ to be constant (17μm), $K$ to be 2.75MPa and $W$ to be 0.19J/m$^2$, the calculated value of $a_o$ using equations 7 and 8 is found to agree rather well with the experimental data (Figure4B). Empirically, it is found that all the $a_o$ vs $P$ data collapses onto a single master curve when the outer radius is divided by $R^{0.5}$ (*See Appendix A1*).

Instead of detaching the lens quasistatically, if it is released at a finite vertical speed (~1.25mm/s), the viscous nectar patch does not relax immediately. Instead, a patch of the solution of radius equivalent to that of the outer ring of contact ($a_o$) is separated from the nectar coated glass slide, which subsequently relaxes and forms a hemisphere on the PDMS lens (Figure5A).

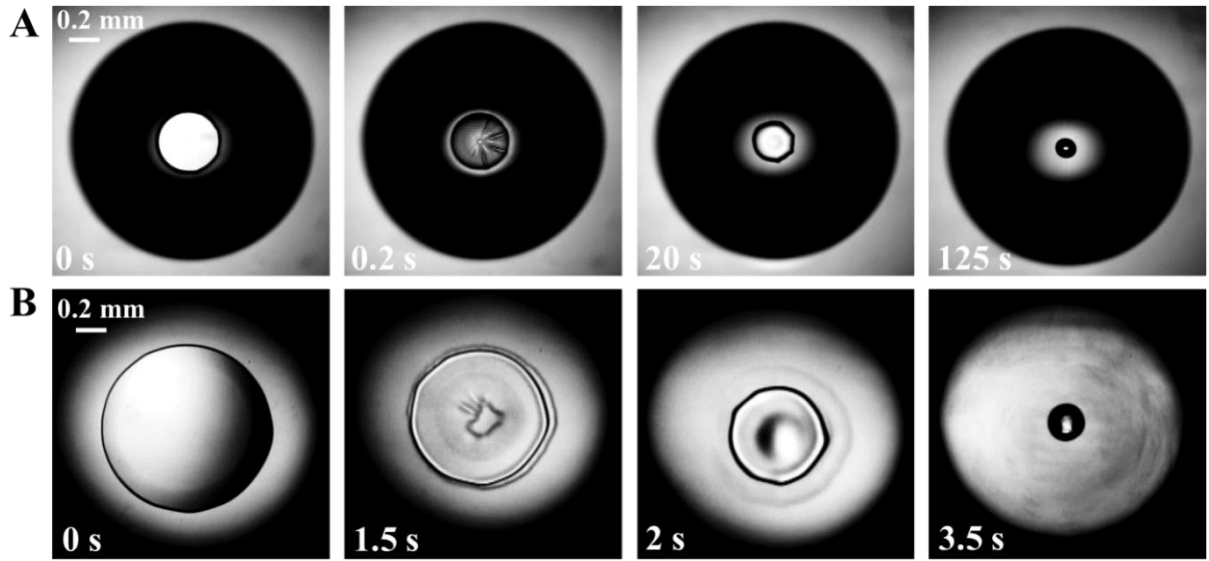

**Figure 5.** (A) A PDMS lens (radius 0.76mm) brought in contact with a syrup coated glass slide and its consequent separation leads to fracture of the syrup film that leaves a tiny amount of syrup on the lens which retracts to become a perfectly hemispherical droplet (B) Another PDMS lens (radius 1.26mm) with a syrup droplet pressed against a PS Petri dish and its consequent separation from the petri dish leaves a small amount of syrup on the Petri dish that contracts into a hemispherical drop.

Similarly, when this drop is brought against a hydrophobic polystyrene Petri dish by pressing the PDMS lens in contact with the Petri dish, it flattens. Upon the rapid release of the lens from the Petri dish, the circular patch shrinks from the latter by a dewetting process. What is intriguing during the dewetting process is that a bump suddenly appears inside the film, which evolves to a hemispherical drop (Figure5B). In the entire process, almost half of the drop is transferred from PDMS lens to the polystyrene substrate, which eventually dries out. The volume of the daughter lens would be proportional to $\rho^3 f(\theta)$, where $\rho$ is its radius and $\theta$ is its contact angle. From the consevation of volume $\rho^3 f(\theta)$ would be same as the volume of the liquid patch of radius $a_o$ that was first separated from the sugar solution coated glass slide. During the final stage of the separation of the lens, we expect a pinch-off instability, in which the surface of the patch relaxes from a deflected state. If the patch begins to dewet from a slightly convex state with a local contact angle $\phi$, the volume of this patch would be ~ $a_o^3 f(\phi)$. Thus, it is plausible that it is the lateral size of the liquid patch which determines its diameter when it fully retracts. In other words, $\rho \sim a_o \sim R^{0.5}$, which is close to the observed relationship: $\rho \sim R^{0.6}$ (Figure3D). A detailed understanding of this problem requires a dynamic analysis considering the dewetting kinetics and the associated hydrodynamics, which is beyond the scope of this paper.

**6. Formation of an array of super-hemispherical caps**

Upon contacting a PDMS block endowed with an array of convex protrusions (i.e, replica of a micro-milled Plexiglass) (Figure6C) against a spin-cast film of aqueous sugar solution on glass, microscopic sugar droplets were transferred to all the protrusions (*See Section 9.3 and 9.4, Experimental details*).An array of hygroscopic precursors could also be transferred to a polystyrene Petri dish, by stamping the above block against it (Figure6D). Upon exposure to water vapor, these hygroscopic precursors transformed to the desired array of super-hemispherical droplets of sugar solution (Figure6E). Using the standard techniques of soft lithography, first a negative (Figure6F) and then a positive image of these droplets were permanently transferred onto a cross-linked PDMS elastomer (Figure7).

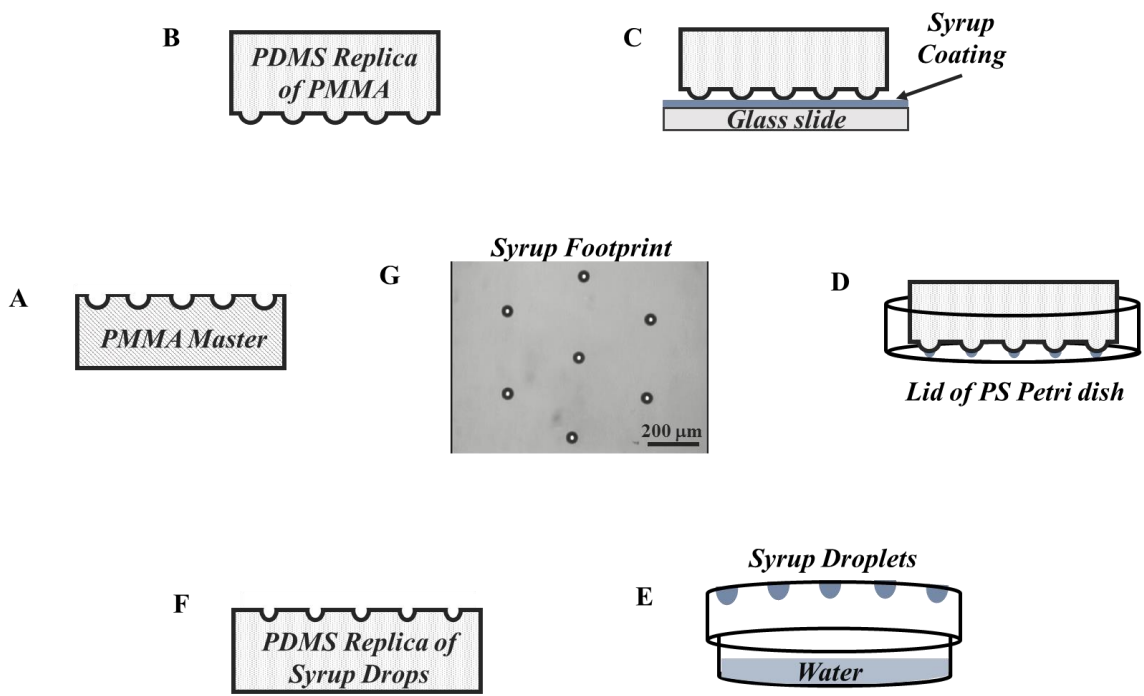

**Figure 6.** (A) A rigid Plexiglass block with an ordered array of concave depressions micromilled onto its surface. (B) A PDMS replica of the Plexiglass block. (C) Contact and separation of the PDMS master against a sugar solution coated glass slide results in the transfer of micro-droplets on the PDMS master (D) Transfer of the micro-droplets onto the lid of a polystyrene Petri dish. (E) Growth of the hygroscopic precursors of sugar solution in a humid atmosphere. (F) Curing PDMS elastomer against the sugar micro-droplets creates concave depressions. (G) Shows the footprint of the syrup drops deposited by the stamp on the polystyrene Petri dish.

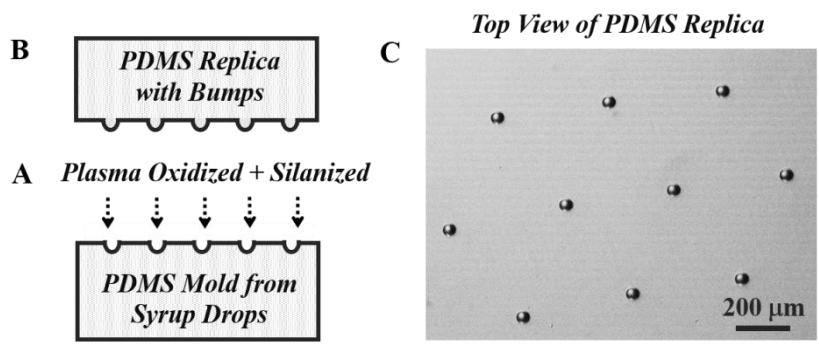

**Figure 7.** (A) Following plasma oxidation and silanization of the PDMS mold of (Figure 4), a negative image of it was produced in another PDMS elastomer (B) in the form of super-hemispherical caps. (C) An optical micrograph of the PDMS replica obtained in (B).

As Figure7 shows, the arrangement of the original array of the microscopic protrusions is faithfully replicated in the final mold, although the ultimate size of the super-hemispherical cap is about $1/4^{th}$ times that of the original one. It is also gratifying to note that the relationship between the base radius of the final super-hemispherical cap and the radius of the starting lens is the same as that reported in Figure3D. An additional important point is that the final micro-asperities generated in this way exhibit a much smoother topology compared to that of the original stamp (Figure8). When the array of lenses, prepared as above, are pressed against a silanized glass slide under a compressive load of 2.8mN per lens, a smooth unbroken circular contact line develops in the zone of contact. By contrast, the PDMS parent lenses that were replicated from the micro-milled Plexiglass exhibited uneven broken contacts when they were pressed against the same smooth substrate. This observation clearly testifies that the daughter lenses prepared as above could serve as smooth model asperities that are indeed suitable for contact mechanics experiments. At present, we have not attempted to analyze the observed deformations for the smoothlenses as the strain was large enough that

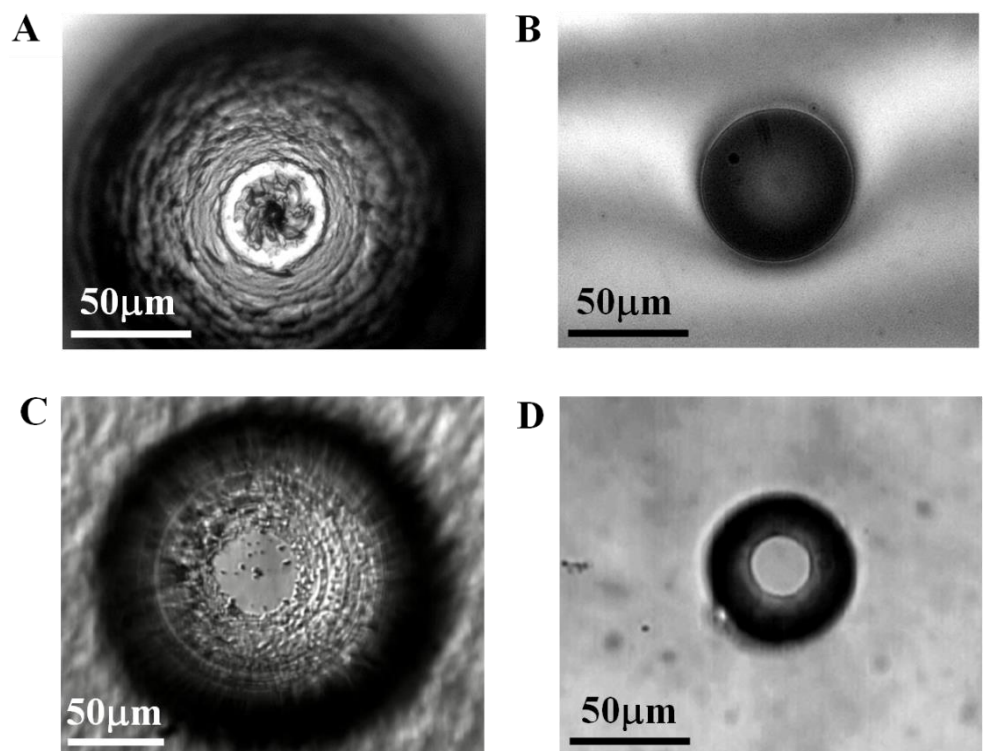

**Figure 8.** Optical micrograph of a parent rough PDMS lens (A) contrasts that of a daughter lens (B). The rough contact (C) of the PDMS lens shown in (A) against a smooth silanized glass slide. This contrasts the smooth contact (D) as the daughter lens is pressed against the same glass slide.

the lenses underwent a non-linear elastic deformation, which is beyond the standard JKR theory.[30]

The final micro-asperities are also quite uniform in size. Analysis based on more than 600 hemispheres yielded radii of curvature of the replicated micro lenses to be $25\pm1.3\mu m$ starting with hemispherical caps of radius $99\pm1\mu m$, thus confirming that this method offers great degree of fidelity in the replication process. Analysis of the heights and base radii of the micro bumps, as obtained from profilometric data, yielded a contact angle of 120°, which is in compliance with that (~130°) acquired from optical contact angle measurements.

**7. Generation of complex hierarchical structures**

In this section, we outline that an additional level of complexity can also be introduced by generating a secondary level of structures on the first generation of hemispherical caps. This can be achieved by coating a micro-fibrillar PDMS (Figure9) with aqueous sugar droplets and transferring them to the pre-fabricated smooth PDMS lenses using the method of stamping as discussed above. The resultant surface morphology can then be cloned further by crosslinking liquid PDMS against it and carrying out the subsequent lithographic processes as discussed previously. A magnified view of such secondary structures supported on the first generation of hemispherical lenses is shown in Figure9I. It is noteworthy that the syrup droplets obtained by stamping of the flat ended-fibrils have sizes similar to those of the fibrils.

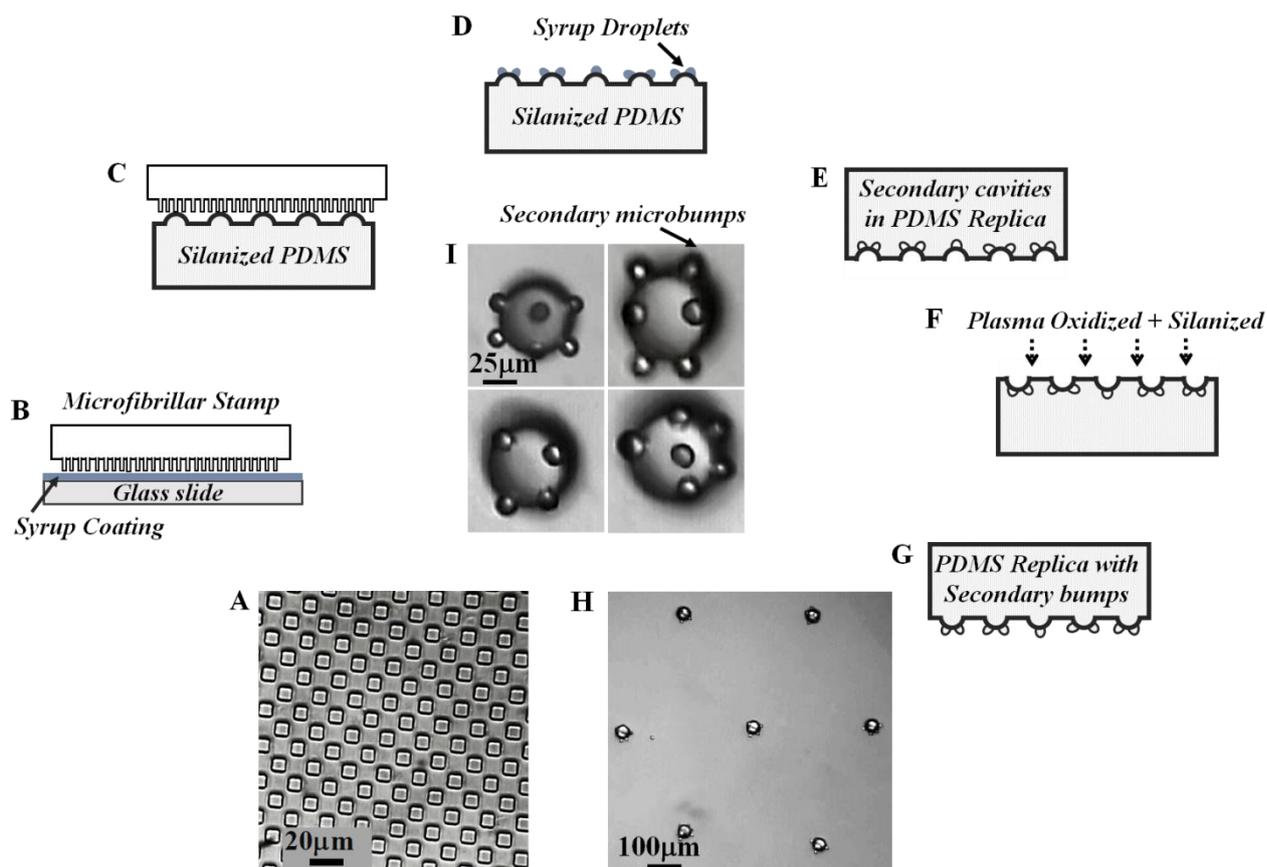

**Figure 9.** Schematic of the preparation of second level structures on the PDMS super-hemispherical caps described in (Figure 5). (A)A micro-fibrillar PDMS (B) is brought into contact with a glass slide spin-coated with a thin layer of syrup. (C) This thin fibrillar PDMS was used to transfer micro-droplets on the super-hemispherical caps of a silanized PDMS block. Crosslinkable PDMS is poured over the sample containing the secondary micro-drops shown in (D) and cured to generate its replica that has secondary cavities as shown in (E). (F) The sample obtained in (E) is plasma oxidized followed by its silanization and used as a mold to generate its replica as can be seen in (G). (H) A transmission optical micrograph showing the secondary structures on the PDMS super-hemispherical caps described in (G). (I) A magnified transmission optical micrograph of four such PDMS super-hemispherical caps bearing secondary structures.

## 8. Re-iterating the main points

In this study we described a new method of manufacturing an ordered array of smooth super-hemispherical caps starting with a rigid Plexiglass substrate possessing concave depressions that, to begin with, may have microscopic and/or sub-microscopic corrugated textures. In the ultra-soft lithographic process as described here, a concentrated aqueous solution of sugar (stabilized by glucose) plays a very important role in providing the nucleation sites for water absorption and growth. The advantage of using a concentrated sugar solution is twofold:

firstly it enhances the growth of the drops without any surface nucleation in the intervening space, and secondly the final drop size can be controlled by manipulating the exposure time to water vapor. Therefore, from one original 'master', it should be possible to make hierarchical levels of super-hemispherical caps of different sizes by regulating the exposure time of the nucleation sites to water vapor, thereby controlling the size of the sugar droplets. We demonstrated that additional complexity can also be introduced onto these primary structures by generating higher level (secondary and even tertiary) patterns using a pre-existing morphology. The resultant structures may find interesting applications in the fabrication of compound lenses in addition to serving as model asperities in contact mechanics, friction and super-wettability[33] studies.

What we have not discussed in the text is that a random distribution of super-hemispherical caps can also be produced on a surface, without having to resort to a stamping process, simply by sprinkling finely powdered sugar on a hydrophobic surface, and exposing it to water vapor saturated air. Upon absorption of water, each of these particles can turn into a super-hemispherical drop of sugar solution that results in a random distribution of hygroscopic droplets on the surface. These droplets can subsequently be cloned into a PDMS elastomer by carrying out the usual crosslinking of PDMS against them and then carrying out the subsequent replication processes as described in this paper. A specific type of random distribution (e.g. Gaussian or non-Gaussian) of such super-hemispherical caps can also be fabricated starting with a Plexiglass substrate having a pre-determined random distribution of concave cavities and carrying out the subsequent replication processes as described here. One point to be noted here is that the size of the final smooth lenses is about 25% of that of the parent lenses. While such a size reduction could be advantageous in certain applications, it is also entirely possible to produce lenses (the second generation super-hemispherical caps as shown in (Figure9I), the final sizes of which will be same as that of the original if a fibrillar stamp with flat termini is used (Figure9A). The studies reported here raise some fundamental issues that could lead to the developments of interesting science, some of which are highlighted below.

The issues of contact mechanics underlying the transfer of the sugar patch to a PDMS lens and its evolution to a spherical cap are profoundly interesting from the point of view of contact mechanics as coupled to capillarity and hydrodynamics. The fact that the inner contact of the PDMS against the sugar solution coated glass follows the contact mechanics theory of Johnson, Kendall and Roberts, where the adhesion is governed by the capillarity in the open surface of the crack is, as far as we know, a new development. The elastic stress generated in the contact region must be supported by the pressure originating from the lubricating flow of the thin film sandwiched between the lens and the glass substrate. Apparently, what has helped us in analyzing the problem using standard contact mechanics is that the high viscous stress slows down the hydrodynamic flow and the evolution of the contact so that the experiments could be performed in a quasistatic manner. Further elaboration of this problem promises to be an interesting study of elasto-capillary driven hydrodynamics that could not only shed more light in the current problem, but it could have implications in soft lubrication and flexographic processes. Fortunately, there are precedences[34-39] of such an analysis based on which more illustrative studies can be perfomed.

Although concentrated sugar solutions were used to carry out the current ultra-soft lithographic processes, other types of hygroscopic solutions, such as the commonly available natural plant nectars can also be used in these studies. Kinetics of the growth of both the sugar droplet, as well as the dry band around it, promises to furnish important information about the chemical potential of water that is modified by its interaction with various sugars. Carrying out similar studies using the solutions[40] of salts taken from a Hofmeister series are expected to shed additional insights into the mechanism of the growth of both the droplet as well as the dry band around it in a humid atmosphere.

**9. Experimental Details**

*9.1. Materials:* The concentrated sugar solution was prepared by replicating as closely as possible the chemical composition of Agave nectar[41] by adding 54% (w/w) of D-Fructose (Fisher Science Education) and 18% of (w/w) D-Glucose (Fisher Chemicals) in deionized (DI, Thermo Scientific Barnstead E-pure unit) water. The viscosity of this solution at room temperature (21°C), as measured using Ostwald viscometer, was found to be 13cp. The mixture was heated to 65°C with constant stirring till a homogeneous solution was obtained. All the replication processes described in this work were carried out with the widely used Polydimethylsiloxane (PDMS) (Sylgard 184, Dow Corning). A Plexiglass block engraved with hemispherical concave cavities micro-milled onto its surface using a desktop CNC Mini-Mill machine (Minitech Machinery Corp., USA) was used as the original master. When it was needed to render a surface hydrophobic, it was reacted with 1*H*, 1*H*, 2*H*, 2*H* Perfluorooctyltrichlorosilane (97%, Lancaster Synthesis Inc.) in the vapor phase using a previously reported[42] procedure. In the last stage of our work, a lithographically produced micro-fibrillar PDMS[43] endowed with fibrils having a square cross-section of 10μm and inter-fibrillar spacing of 15 μm was used for the generation of complex elastomeric structures. All phenomena were imaged using a CCD camera (Sony XC-75 or MTI-72) in a Nikon Diaphot microscope and subsequently analyzed using image software Virtual Dub and ImageJ.

*9.2. Preparation of a single super-hemispherical lens:* Model asperities were made in the form of single elastomeric lenses ranging in size from 0.1mm to 1.3mm radii of curvature. In order to achieve this, syrup droplets were deposited on the inner side of the lid of a polystyrene Petri dish and allowed to grow in size by exposure to water vapor for 5 minutes. At this stage, a degassed mixture of PDMS was poured onto the lid, followed by curing in a hot air oven at a temperature of 75°C for 2 hours to generate a replica of the drops in the form of concave super-hemispherical cavities. (Note: all the structures having contact angle greater than 90° are referred to as super-hemispheres in the text). The replicas were then oxidized (Harrick Plasma), silanized by reacting it with Perfluorooctyltrichlorosilane and cloned onto PDMS in the form of super-hemispherical lenses of varying sizes.

*9.3. Stamping of syrup droplets:* A Plexiglass block was micro-milled to produce 675 hemispherical concave cavities engraved on its surface in an ordered hexagonal packing with inter cavity separation of 216μm, with a ball end mill of radius 100μm. The concave cavities in the Plexiglass thus had a radius of curvature of 100μm and were almost hemispherical with contact angles little less than 90°. A degassed mixture of the crosslinkable liquid PDMS was first poured onto the Plexiglass block and cured in the oven at 75°C for 1.5 hours in order to obtain the stamp used to generate the ordered pattern of smooth elastomeric lenses. The convex hemispherical asperities were however rather rough at microscopic length scales as can be discerned even with a low power optical microscope. This replica, endowed with uniformly distributed convex hemispherical asperities, was next brought into contact with a spin-cast film of aqueous sugar solution on a glass slide and subsequently separated that resulted in the transfer of small liquid droplets at the tips of the hemispheres. In order to obtain reproducibility of the transfer process, the following method was used. The flat side of the stamp was affixed to a lightweight circular glass disc (diameter 25mm, 9 grams) that hung from the end of a metallic chain (length 45mm), which is stiff under tension, but flexible under compression. Care was taken to make sure that the circular disk was parallel to the sugar solution coated glass slide as best as possible via visual inspection. The disc itself could be translated in three orthogonal directions by means of an XYZ micromanipulator (Nikon Narishige). Before carrying out the stamping process, the PDMS stamp was thoroughly cleaned with warm water followed by DI water, and then dried by blowing ultrahigh purity Nitrogen gas over it.

The aqueous sugar solution was spun over a clean glass slide (75mm x 25mm x 1mm) using a Photo-Resist Spinner (Headway Research, Inc.) at a speed of 2500rpm for 20 seconds that resulted in a film of thickness $18\pm1$μm. Selection of the speed and time of spin-coating had to be adjusted judiciously depending on the viscosity of the solution being used. A thin yet sticky film was necessary for the transfer of the aqueous sugar drops onto the spherical asperities of the stamp and its subsequent transfer to a hydrophobic polystyrene Petri dish. After fixing the sugar solution coated glass slide onto a sample holder stage, the PDMS stamp was brought into contact with the slide using the micromanipulator (Figure10).

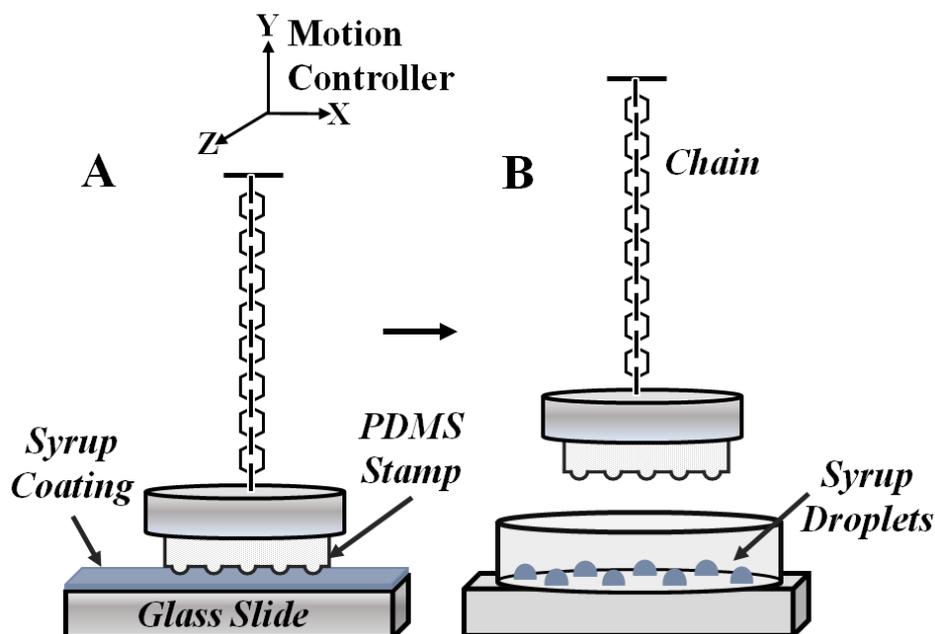

**Figure 10.** Schematic of the setup used for the stamping of aqueous sugar droplets onto a hydrophobic surface. (A) The PDMS stamp supported on a glass disk is brought into contact (A) and then separated. (B) Upon contact with and then separation of this PDMS stamp containing the sugar droplets against a hydrophobic surface results in the hygroscopic precursors for water vapor condensation. Please note that the schematics are not drawn to scale.

Immediately following the aforementioned contact, the stamp was separated from the glass slide. This PDMS block was then stamped against a clean polystyrene Petri dish following contact and separation using the same micromanipulator. All these processes were carried out atop a vibration isolation table (Micro-G, TMC) to dampen the effects of ground noise. Optical microscopic observation of the stamped Petri dish verified the presence of an ordered pattern of droplets of sugar solution having uniform sizes. After these drops dried in air for about 30 minutes, they were exposed to a water saturated atmosphere for 5 minutes, thus allowing them to grow by absorption of water vapor. At this stage, a degassed a mixture of Sylgard 184 was gently poured inside the Petri dish fully covering the aqueous sugar droplets and cured at 75°C for 1.5 hours. This generated a replica of the ordered pattern of super-hemispherical droplets in the form of concave depressions on the surface of the cured PDMS block. Since sugar is highly hygroscopic, a concentrated solution of it retains water with much more tenacity than pure water drops, thus restricting its evaporation while curing at high temperature.

*9.4. Formation of an array of super-hemispheres:* After appropriate cleaning, the replica of the PDMS master endowed with concave super-hemispherical depressions was plasma oxidized following which it was silanized by reacting it with Perfluorooctyltrichlorosilane by flushing the vapor of the silane with purified dry air as a carrier over the substrate. Following the silanization of the PDMS mold for 5 minutes, a degassed mixture of Sylgard 184 was poured over it and cured at 75°C for 1 hour. Upon curing, the PDMS replica with ordered array of convex asperities could be easily peeled off the mold without any damage to the former. These PDMS blocks were cleaned thoroughly with DI water, dried by blowing ultrahigh purity Nitrogen gas and then stored inside clean Petri dishes.

*9.5. Contact Mechanics Related Studies:* All the loading and unloading experiments involving the PDMS hemispheres were carried out using the standard equipments and procedures under ambient conditions (21°C, 35% relative humidity) that are well-documented in the literature. All the details can be found in reference 30.

**Appendix**

**A1. Collapse of outer radii of the JKR contact by normalizing them with the radii of the lenses**

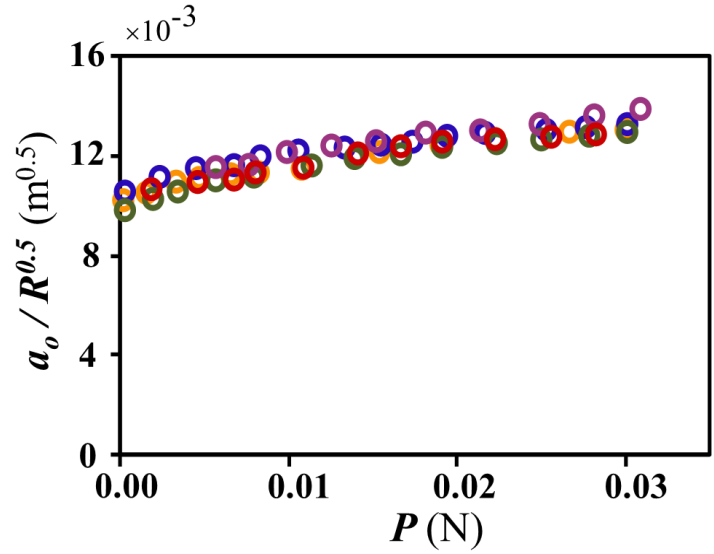

**Figure A1.** This figure illustrates that when the outer radius ($a_o$) of Figure 4B is divided by $R^{0.5}$, all the data collapse onto a single plot.

**A2. Kinetics of growth and decay of syrup drop.**

The kinetics of the growth of a syrup droplet, where the driving force arises from the difference of the partial pressure of the vapor ($P^*$) far from the drop and that ($P$) just outside of it is as follows:

$$\frac{dN_w}{dt} = kA[P^* - P] \tag{9}$$

$$A = (36\pi)^{1/3} V^{2/3} \tag{10}$$

Where, $A$ and $V$ are respectively the area and the volume of the drop. Combining equations 9 and 10 we have:

$$\frac{dN_w}{dt} = (36\pi)^{1/3} k[P^* V^{2/3} - P V^{2/3}] \tag{11}$$

Where $P = HC_w$ and $C_w = N_w/V$. Thus, equation 11 can be written as:

$$\frac{dN_w}{dt} = (36\pi)^{1/3} k [P^* V^{2/3} - H N_w V^{-1/3}] \tag{12}$$

Where the total volume of the drop is: $V = N_w \bar{v}_w + N_s \bar{v}_s$

Replacing the value of $V$ in equation 12 we obtain the following:

$$\frac{dN_w}{dt} = (36\pi)^{1/3} k [P^* (N_w \bar{v}_w + N_s \bar{v}_s)^{2/3} - H N_w (N_w \bar{v}_w + N_s \bar{v}_s)^{-1/3}] \tag{13}$$

For the desorption phase of the droplet, the governing kinetic equation is:

$$-\frac{dN_w}{dt} = kAP \tag{14}$$

Expressing $P$ in terms of $H$, $N_w$ and $V$ (as has already been shown above) results:

$$-\frac{dN_w}{dt} = k H N_w (N_w \bar{v}_w + N_s \bar{v}_s)^{-1/3} \tag{15}$$